\documentclass[12pt,preprint]{aastex}
\usepackage{rotating}
\usepackage{amsmath}
\newcommand \kms{km~$\rm{s}^{-1}$}

\newcommand \msol{M$_{\odot}$}

\newfont{\rten}{cmr10} 
\def \arcdeg{\hbox{$^\circ$}}

\begin{document}

\title{Will the Large Synoptic Survey Telescope detect extra-solar planetesimals entering the solar system?}

\author{Amaya Moro-Mart\'{\i}n$^{1,2}$}
\author{Edwin L. Turner$^{3,4}$}
\author{Abraham Loeb$^{5}$}

\altaffiltext{1}{Centro de Astrobiolog\'{\i}a (CSIC-INTA), 28850 Torrej\'on de Ardoz, Madrid, Spain}
\altaffiltext{2}{Department of Astrophysical Sciences, Princeton University, Princeton, NJ 08544, USA}
\altaffiltext{3}{Princeton University Observatory, Princeton, NJ 08544, USA}
\altaffiltext{4}{Institute for the Physics and Mathematics of the Universe, University of Tokyo, Kashiwa, Chiba 277-8568, Japan}
\altaffiltext{5}{Harvard University, Center for Astrophysics, MS 51, 60 Garden Street, Cambridge MA 02138, USA}

\begin{abstract}
Planetesimal formation is a common by-product of the star formation process.  Taking the dynamical history of the solar system as a guideline -- in which the planetesimal belts were heavily depleted due to gravitational perturbation with the giant planets -- and assuming similar processes have take place in other planetary systems, one would expect the interstellar space to be filled with extra-solar planetesimals. However, not a single one of these objects has been detected so far entering the solar system, even though it would clearly be distinguishable from a solar system comet due to its highly hyperbolic orbit. The {\it Large Synoptic Survey Telescope} will provide wide coverage maps of the sky to a very high sensitivity, ideal to detect moving objects like comets, both active and inactive. In anticipation of these observations,  we estimate how many $inactive$ "interstellar comets" might be detected during the duration of the survey.  The calculation takes into account estimates (from observations and models) of  the number density of stars, the amount of solids available to form planetesimals, the frequency of planet and planetesimal formation, the efficiency of planetesimal ejection, and the possible size distribution of these small bodies. 
 \end{abstract}

\section{INTRODUCTION}
The study of protoplanetary and debris disks indicates that planetesimal formation is a common by-product of the star formation
process, and there is observational evidence that in some cases it has led to the formation of planets. 
Taking the dynamical history of the solar system as a guideline, in which efficient dynamical ejection events triggered by 
gravitational perturbation with the giant planets were likely responsible for the depletion of the asteroid belt and the 
Kuiper belt (KB), and the formation of the Oort Cloud (OC), and assuming similar processes have taken place in other planetary 
systems, one would expect the interstellar space to be filled with extra-solar planetesimals. However, not a single one of these 
extra-solar planetesimals has been detected so far entering the solar system, even though it would be clearly distinguishable 
from a solar system comet due to its hyperbolic orbit resulting from the velocity of the Sun with respect to the Local Standard of 
Rest (LSR; $v_{LSR}$ = 16.5 \kms). 
McGlynn \& Chapman (1989) addressed this lack of "extra-solar comet" detections by comparing it to the expected 
number of detections based on a simple model that considers that stars have a spatial density of 0.1 pc$^{-3}$, 
each of them ejects $\sim$ 10$^{14}$ planetesimals into interstellar space, and the entire ensemble of planetesimals has 
a velocity distribution with respect to the Sun similar to that of the stars in the local neighborhood. Including 
corrections for the observability of these extra-solar comets, McGlynn \& Chapman (1989) concluded that there is a missing 
extra-solar comet problem because a significant number of extra-solar 
comets should have already been detected. 
However, Jewitt (2003) argued that if each star ejects 10$^{13}$ planetesimals with $r > 1$ km 
(resulting in a number density of planetesimals of 10$^{-3}$ AU$^{-3}$)
and a velocity relative to the Sun of 20 \kms, one would expect 0.3 planetesimals per year to cross within 5 AU of the Sun (5 AU being the detection distance 
with Pan-STARRS for a low albedo comet-like body).  Because a 1 km size object at 5 AU corresponds to about 24 mag, Jewitt (2003) concluded 
that it is no surprise none have been found so far and that  Pan-STARSS would either detect a few interstellar comets over its 10 year lifetime or place limits on their number density. 

The last few years have witnessed a revolution in the fields planetary system studies (solar and 
extra-solar) with the completion of extended surveys of protoplanetary and debris disks, and most
importantly, with the discovery and characterization of extra-solar planets and of Kuiper Belt objects (KBOs), shedding new 
light into the amount of solid material that might be available to form planetesimals,  the frequency of planetesimal and 
planet formation, and the size distribution of small solar system bodies, respectively.  These are the key features that 
affect the model calculations in McGlynn \& Chapman (1989) and Jewitt (2003) in a fundamental way. In the future, cometary observations will
witness a revolution on its own when Pan-STARRS and the  {\it Large Synoptic Survey Telescope} (LSST) become operative, as they will provide wide coverage maps of the sky
to a very high sensitivity ideal to detect moving objects like comets (active and inactive).
Because of these new developments, there is the need to revisit the interstellar comet estimates.

To calculate the number density of planetesimals in interstellar space, McGlynn \& Chapman (1989) and Jewitt (2003) assumed a number density of stars that is independent of stellar mass, a fixed number of planetesimals each star contributes (also independent of stellar mass), and that all planetesimals are km-sized. We improve these approximations by taking into account the following observational and modeling results.
\begin{itemize}
\item The {\it number density of stars} depends on the stellar mass (Section \ref{stellar_density}). 
\item {\it Protoplanetary disks} studies indicate that the amount of solids available to form planetesimals also depends on the stellar mass (Section \ref{solids}). 
\item {\it Debris disks}  surveys confirm that for a fraction of stars this solid material has accreted to form planetesimals. But debris disk surveys are limited by sensitivity to $\sim$100 times the amount of dust in the solar system; this observational limitation, together with theoretical expectations, indicate that it is reasonable to assume all systems harboring giant planets also formed planetesimals (Section \ref{debris}). 
\item Giant planets eject planetesimals into interstellar space very efficiently (Section \ref{ejection}). 
\item {\it Radial velocity surveys} indicate that a fraction of stars harbor giant planets. Because the presence of massive planets seems necessary to provide a mechanism to scatter planetesimals into interstellar space and, as we mentioned above, it is reasonable to assume all systems harboring giant planets also formed planetesimals, we will assume that the stars contributing to the population of planetesimals in interstellar space are those that harbor giant planets (Section \ref{planets}). 
\item Not all planetesimals ejected are km-sized. We estimate the number density of planetesimals with $r>R$ based on the observed size distribution of the small body population in the solar system and on theoretical considerations (Section \ref{size_dist}). 
\item Assuming the  planetesimals are distributed isotropically in interstellar space and that they have an albedo of 6\% (similar to that of inactive comets), we calculate the luminosity function of extra-solar planetesimals as observed from the Earth (Section \ref{luminosity}). 
\item Finally, using the luminosity function above and simple considerations about the characteristics of the LSST survey,  we estimate the expected number of "interstellar comet" detections (Section \ref{dicussion}). 
\end{itemize}

\section{MASS DENSITY OF EXTRA-SOLAR MATERIAL}
\subsection{Number Density of Stars}
\label{stellar_density} 
We adopt the stellar density calculated by Kroupa, Tout \& Gilmore (1993): the number  density of  stars per pc$^3$, out to $\sim$ 
130 pc from the Sun, within the midplane of the galaxy, and with 
stellar masses between {\it M$_{*}$} and {\it M$_{*}$+dM$_{*}$} (in units of M$_{\odot}$), is given by 
\begin{equation}
\begin{split}
n(M_*) = \xi(M_*)dM_*~~~~{\rm with,}~~~~~~~~~~~~\\ 
\xi(M_*) = 0.035 M_*^{-1.3}~~~~{\rm if~0.08} \le M_* < 0.5\\
\xi(M_*) = 0.019 M_*^{-2.2}~~~~~{\rm if~0.5} \le M_* < 1.0\\
\xi(M_*) = 0.019 M_*^{-2.7}~~~~{\rm if~1.0} \le M_* < 100.  
\end{split}
\label{eq_stellar_density}
\end{equation}
The binary fraction of stars is approximately 50\%. 

\subsection{Total Mass Available to form Solids per Star}
\label{solids}
The contribution of each star to the number of planetesimals in interstellar space is going to depend on the amount of solid material (primordial dust) that surrounds each star and that might be available to accrete into larger bodies. Andrews \& Williams (2007) conduced a submillimeter survey of 170 Class II sources (pre-main-sequence stars with evidence of thick protoplanetary disks) in two star-forming regions, Taurus  ($<$age$>\sim$1 Myr) and $\rho$-Oph  ($<$age$>\sim$0.7 Myr). It was found that the disk mass  distributions are very similar with $<$M$_{disk}> \sim$ 0.005 \msol~and $<$M$_{disk}/M_*>\sim$ 0.01 (where $M_{disk}$ includes both gas and dust),  
for stellar masses ranging from 0.1 to 3 \msol~(peaking at 0.4 \msol), and assuming a disk size of 100 AU and a gas-to-dust ratio of 100:1 (Andrews \& Williams 2007). This is comparable to the minimum-mass solar nebula ($\sim$0.015 M$_{\odot}$ -- including gas and dust), which is the total disk mass  required to account
for the condensed material in the solar system planets. 
Even though $M_{disk}/M_*$ decreases with stellar age from a few percent at 0.1 Myr to 30$\times$ smaller value at 10 Myr, these Class II disk masses indicate that  $\sim 10^{-4}M_*$ of solids could have been available to form planetesimals. Andrews \& Williams (2007) note that their disk mass estimates adopted an opacity characteristic of mm-sized bodies.  If larger bodies were to be present (cm or meter-sized), the opacity would be smaller than assumed and therefore the disk mass would be underestimated by up to an order of magnitude. Because the presence of these larger grains is currently unconstrained by the observations (except in the case of TW Hydra -- Wilner et al. 2005), in this paper we will assume $M_{solids} \sim 10^{-4}M_*$.  

\subsection{Fraction of Stars that Formed Planetesimals}
\label{debris}

The disks of dust observed\footnote{Most of the 300 debris disks known to date are spatially unresolved and were identified from the presence of an infrared excess in the spectral energy distribution, attributed to the thermal emission from dust particles heated by the central star.} around mature main-sequence stars of a wide range of spectral types indicate that, in some cases, there is evidence that the primordial dust particles in the protoplanetary disks have  accreted to form planetesimals. This is inferred from the fact that the lifetime of the dust particles in these dust disk -- under the effect of mutual collisions, Poynting--Robertson drag and radiation pressure -- is 10$^{4}$--10$^{6}$ years, much shorter than the age of the star, 10$^{7}$--10$^{10}$ years, implying that the dust observed in these mature systems is not  primordial, but must be replenished by a reservoir of dust-producing planetesimals. This is why these dust disks are known as debris disks. The characteristic dust temperatures inferred from the spectral energy distributions, together with the images of the two dozen of spatially resolved debris disks known to date,  indicate that the debris disks have sizes comparable to the distribution of the small bodies in the solar system (from tens to few hundreds of AU). 

The {\it Spitzer Space Telescope} carried out extensive surveys to characterize the frequency and properties of the debris disks around more than 700 stars of different spectral types, ages and environment. For the purposes of this paper, we are interested in  the fraction of stars that show evidence of dust excess emission, i.e., the fraction of stars that show evidence of harboring (dust-producing) planetesimal. 

\begin{itemize}
\item {\it solar-type stars}: for stars younger than 300 Myr, the disk frequency at 24 $\mu$m -- tracing dust in the asteroid belt-like region -- is 14.7\%, decreasing to  2.7\% for older stars; at 70 $\mu$m -- tracing dust in the KB-like region -- the disk frequency is 6\%--10\% and does not change significantly with age (from a survey of 328 F5--K7 stars --  
Hillenbrand et al. 2008; Meyer et al. 2008; Carpenter et al. 2009).  
A different survey of 225 F0--G9 stars older than 600 Myr yield disk frequencies of  4.2$^{+2}_{-1.2}$\% at 24 $\mu$m  and 16.4$^{+2.8}_{-2.9}$\% at 70 $\mu$m (Trilling et al. 2008). 
\item {\it A-type stars}: for stars younger than 30 Myr, the disk frequencies are 42\% at 24 $\mu$m -- tracing dust at $\leq$5--50 AU -- and 48\% at 70 $\mu$m -- tracing dust at 50--200 AU; for stars older than 400 Myr, these numbers decrease to 2\% at 24 $\mu$m and 12\%  at 70 $\mu$m (from a survey of 160 B9--A7 single stars -- Su et al. 2006; Rieke et al. 2005). 
\item {\it Low-mass stars}: the fraction of stars with debris disks  decreases steeply for stars  later than K2, with an
excess rate for old M stars of 0\% with upper limits (binomial errors) of 2.9\% at 24 $\mu$m and 12\%
at 70 $\mu$m (Gautier et al. 2007; Beichman et al. 2006; Trilling et al. 2008). 
This may be an observational bias  because debris disks around this low-mass stars would have a peak emission 
at $\lambda$ $>$ 70 $\mu$m and therefore remain too cold to be detected by {\it Spitzer}. 
\item {\it Binary stars:} the frequency of debris disks around binary stars   is similar or even higher than around single stars of similar type (from a survey of 69 A3--F8 binary systems -- Trilling et al. 2007). 
\end{itemize}

The above debris disk fractions are lower limits because of the limited  sensitivity of the {\it Spitzer}  observations: 
Bryden et al. (2006) found that the frequency of dust detection 
increases steeply as smaller fractional luminosities ($L_{dust}/L_*$) are considered, going from nearly 0\% for $L_{dust}/L_*$ =  10$^{-3}$, to 
2$\pm$2\% for $L_{dust}/L_*$ =  10$^{-4}$~and 13$\pm$5\% for $L_{dust}/L_*$ =  10$^{-5}$; using the cumulative distribution
of fractional luminosities and assuming that the distribution is a Gaussian, they concluded that 
the luminosity of the solar system dust (with $L_{dust}/L_*$ =  10$^{-7}$ to 10$^{-6}$)  is consistent with being 10 $\times$ brighter or fainter than 
an average solar-type star, i.e., even though no solar system debris disks analog have been found due to limited sensitivity 
of the detectors, the observations indicate that debris disks at the solar system level might be common. 

Other observational evidence also indicate that planetesimal formation is a robust process that can take place under a wide range of conditions: debris disks are present around stars with more than 2 orders of magnitude difference in stellar luminosity, also in systems with and without binary companions (as circumstellar or circumbinary disks), and 
around  stars with a wide range of metallicities (unlike planet bearing stars that are strongly correlated with high stellar metallicities).  This indicates that planetary systems harboring dust-producing planetesimals are more common than those with giant planets\footnote{This is also in agreement with the lack of correlation between the presence of debris disks and the presence of planets (Moro-Mart\'{\i}n et al. 2007)}, which would be in agreement with the core accretion models of planet formation where the planetesimals are the building blocks of planets and the conditions required for to form planetesimals are less restricted than those to form gas giants. 

Because of the above considerations and because, as we will see below, the  mechanism of planetesimal ejection into interstellar space likely requires the presence of massive planets, we will assume that  the fraction of stars contributing to the population of interstellar planetesimals  is determined by the fraction of stars harboring massive planets, rather than the fraction of stars with debris disks. 

\subsection{Ejection of Planetesimals into Interstellar Space}
\label{ejection}

The dynamical history of the solar system can be relatively well constrained through models and observations and
its highlights, with regard to the ejection of planetesimals, can be summarized as follows. 
\begin{itemize}
\item Before 10 Myr after the Sun was formed, while the Sun was still embedded in its maternal stellar cluster and 
before the gas in the primordial protoplanetary disk dispersed, Jupiter and Saturn formed and scattered the planetesimals in the 
Jupiter-Saturn region out to large distances. Some of these planetesimals were completely ejected from the system after multiple
close encounters with the planets. Others remained bound because their pericenters were lifted beyond the orbit of Saturn
(avoiding subsequent encounters with the planets) due to gravitational perturbations from the stars and the gas in the stellar cluster. This population of bound planetesimals formed the OC. A model that accounts for the characteristics of the OC  
estimates that  75--85\% of planetesimals in the Jupiter-Saturn region were completely ejected from the system (Brasser, Duncan \& Levison 2006). 

\item Between 10 and 100 Myr,  O'Brien et al. (2007) estimate that the gravitational perturbations from Jupiter and Saturn and the mutual perturbations amongst the largest asteroids depleted the asteroid belt by a factor of $\sim$100, leaving behind an asteroid belt about 10--20$\times$ more massive than today.

\item At $\sim$ 700 Myr, the slow migration of the giant planets (due to the interaction of the planets with a massive 
trans-Neptunian planetesimal disk) forced Jupiter and Saturn to cross a mean motion resonance making their orbits unstable and triggering a quick rearrangement of the planets orbits into their current configuration (Gomes et al. 2005; Morbidelli et al. 2005; 
Tsiganis et al. 2005). During this process, secular resonances swept through the asteroid belt making their orbits unstable, 
scattering some asteroids into the inner solar system (producing the Late Heavy Bombardment - Strom et al. 2005), while ejecting others out of the  system, resulting in a depletion factor of $\sim$10--20 (i.e. depleting 90--95\% of the asteroids that had remained).  
This planet rearrangement also resulted in the sudden outward migration of Neptune and its exterior mean motion resonances, that swept through the KB that as a consequence got heavily depleted. 
\end{itemize}

Overall, for the solar system it is estimated that only a very small fraction of negligible mass of the initial planetesimal disk was left behind. 

As can be inferred from the above description, the efficiency of planetesimal ejection is very sensitive to the dynamical history 
of the system, which in turn depends strongly on the planetary configuration (the number of planets, their semimajor axis, masses 
and mass ratios) and the planetesimal disk (mass, extent and radial profile). The large diversity of planetary systems inferred 
from radial velocity observations warn us that there is no reason to assume that planetary systems share similar dynamical 
histories.  However, the following considerations hint that the efficiency of planetesimal ejection may also be high for many other planetary systems.
\begin{itemize}
\item The high occurrence of Jupiter-sized planets at small semimajor axis indicates that these planets have migrated significantly inward from their formation sites, a migration that likely resulted in the ejection of most of the planetesimals that interacted with the planet. 
\item Highly eccentric planets are also common, however, there is no reason to expect planetary formation processes to yield high 
eccentricities and therefore it has been suggested that they result from periods of dynamical instability in 
multi-planet systems that lead to a sudden rearrangement of the orbits (Ford \& Rasio 2008; Juric \& Tremaine 2008). Models and observations of the solar system indicate that the sudden rearrangement of the giant planets orbits triggered the ejection a large number of planetesimals in the KB and the asteroid belt-regions.
\item Hints of transient events that may be due to dynamical instabilities are found in the observation of several anomalously 
massive debris disks around old stars with inferred levels of dust production that could not be sustained for the age of the system 
(Wyatt et al. 2007). 
%\item Dynamical models show that gravitational scattering can be very efficient; for planets on circular orbits around solar-type stars, the ejection efficiencies are the following  (for test particles that due to migration cross the orbit of the planet): $<$10\% for a 0.3 M$_{Jup}$ mass planet located at 1 AU and  $\sim$20--40\% if located at 5--30 AU; $\sim$60\% for a 1 M$_{Jup}$  mass planet located at 1 AU and $\sim$80\% if located at 5--30 AU; and $\gtrsim$90\% for a 3--10 M$_{Jup}$ mass planet located at 1 AU--30 AU (Moro-Mart\'{\i}n and Malhotra 2005). 
\end{itemize}

For the purposes of this paper, we will assume that as in the solar system,  most of the planetesimals in the planetary systems are ejected and only a small fraction of negligible mass is left behind.

\subsection{Fraction of Stars with Massive Planets}
\label{planets}

Because the mechanism of planetesimal ejection into interstellar space likely requires the presence of massive planets, and as we discussed above, we can assume planetesimals formed in all systems with planets,  the fraction of stars contributing to the population of interstellar planetesimals  is determined by the fraction of stars harboring massive planets.

\begin{itemize}

\item {\it solar-type stars}: radial velocity searches indicate that 6.7\% of solar-type stars have planets with masses $>$ 0.1 M$_J$ and semimajor axis $<$ 5 AU (Marcy et al. 2005); from the extrapolation of these surveys it is expected that  $\sim$19\% harbor giant planets are within 20 AU\footnote{This comes from adopting the following probability for a star to have a planet of mass $M$ at orbital period $P$: d$N$ = $CM^{\alpha}P^{\beta}$dln$M$dln$P$, with $\alpha$ = -0.31$\pm$0.2 and $\beta$ = 0.26$\pm$0.1. Assuming solar-type stars and planet masses 0.3$M_{Jup} < M <$ 15$M_{Jup}$, the frequencies are $\sim$ 8.5\% for $a <$ 3 AU and $\sim$19\% for $a <$ 20 AU (Cumming et al. 2008); if we were to consider larger semimajor axes, the frequencies would be  $\sim$ 23\% for $a <$ 40 AU and $\sim$33\% for $a <$ 100 AU.} (Cumming et al. 2008).

\item {\it A-type stars}: there are no firm statistics because radial velocity  studies are complicated by the rotational broadening of the absorption lines, 
the decreased number of spectral features due to high surface temperature, and a large excess velocity resulting from inhomogeneities and pulsation. 
A way to circumvent  these problems is by studying the cooler intermediate mass subgiants into which  A-type stars evolve.  Johnson et al. (2007a)  found  that 
for 1.3--1.9 \msol~subgiants (corresponding to F and A main-sequence stars) the frequency of planets with masses $>$0.8 M$_{J}$~within 2.5 AU is 8.9$\pm$2.9\% (= 9/101), 
compared to 4.2$\pm$0.7\% (= 34/803) for solar mass stars and 1.8$\pm$1\%  (= 3/169) for  low mass-K and M dwarfs. 
These results indicate that A--F stars are five times more likely than M dwarfs 
to harbor a giant planet, although there is a possible bias from metallicity because the solar-type stars and the 
subgiants in the planet surveys are slightly metal-rich compared to stars in a volume-limited sample.  After correcting for this effect, the above ratio decreases from 5 to 2.5 but is still significant (Johnson et al. 2007a). 

This  positive correlation between the stellar mass and the occurrence of planets is in agreement with core accretion models for planet formation that predict that 
more massive stars result in more massive protoplanetary disks with the increased surface density of solid material in the midplane 
favoring the formation of planetesimals (Ida \& Lin, 2005). Kennedy and Kenyon (2008) predict that the probability that a given star of mass
0.4--3 \msol~harbors at least one giant planet increases linearly with stellar mass. For stars $>$3 \msol~the snow
line quickly moves out to 10--15 AU before protoplanets form, limiting the formation of gas giant planets.  

Here we will assume that the fraction of A-type stars harboring planets is at least as high as for solar-type stars, $\sim$ 20\%. The effect of considering a higher planet fraction can easily be estimated from Table 1. 

\item {\it Low-mass stars}: because $\sim$ 77\%$\pm$10\% of stars are M-dwarfs, it is important to consider their 
possible contribution to the population of interstellar planetesimals. Johnson et al. (2007b) found that out of 300 M dwarfs monitored, 
only two have Jupiter-mass planets, while six have Neptune-Uranus mass planets (giving a total frequency of planets around M-dwarfs of 
$\sim$ 8/300 $\approx$ 2.7\%).  Even though the planet searches around M-dwarfs are limited by small number statistics, low-mass planets  (Neptunes and super-Earths) appear to be relatively more common than Jupiter-mass planets (Bonfils et al. 2007; Endl et al. 2008). 

These observations agree with core accretion models that predict Jupiter-mass planets are relatively rare around M-dwarfs (due to the longer orbital periods, by the time the core reaches the critical mass of 10 M$_{\oplus}$, the gas disk has dissipated and can no longer feed the 
planet's gaseous envelope  -- Laughlin et al. 2004; Ida and Lin 2005; Kennedy \& Kenyon 2008), while Neptune-mass planets form more easily (Wetherill 1996, Laughlin et al. 2004; Ida \& Lin 2005 -- the latter predicting a higher frequency of short-period ice giants around M-dwarfs compared to solar-type stars).  

For the purposes of this paper, we will assume that the fraction of M dwarfs with 0.1 \msol $<$ {\it M} $<$ 0.6 \msol~that can contribute to the population of interstellar planetesimals is $\sim$ 3\%, and that because of the lower gravitational potential of M dwarfs, Neptune-mass planets can efficiently eject planetesimals. 

\item {\it Binary stars}: the dependency of the frequency of planets on the presence of stellar companions is still under investigation.
Bonavita and Desidera (2007) carried out a literature search of companions to the stars in  Fischer and Valenti (2005) sample, with a uniform planet detectability for periods $<$ 4 years. They found that: 
(1) the global frequency of planets in the binary sample (15/202 = 7.4\%) is not statistically different from that of planets in the single star sample (5.3\%). 
Even when making the conservative assumption  that all stars without identified companions in the literature are single, the frequency of planets in 
binaries is no more than a factor  of 3 lower than that in single stars.  (2) The frequency does not depend on the binary 
separation, except for close binaries. (3) For close binaries, the frequency of planets is lower than for single stars and wide binaries (however, 
the RV surveys are biased against  binaries closer than $\sim$2''). On going surveys, consisting on a radial velocity search for planets around spectroscopic binaries and an adaptive optics search for stellar companions to stars with and without planets,   will soon shed light on the dependency of 
planets on binarity. 

In this paper, we will assume that only single stars 
contribute to the population of interstellar comets, adopting a binary fraction of 50\%. 
\end{itemize}

\subsection{Expected Mass Density of Extra-Solar Planetesimals}
\label{mass_density}
With the considerations discussed above, we can now calculate the contribution of stars of masses between 
$M_*$ and $M_*$+d$M_*$ to the total mass density of  extra-solar planetesimals. We assume that
(1) the number density of stars per pc$^3$ is given by Equation \eqref{eq_stellar_density}. 
(2) Each star harbored a protoplanetary disk with a total mass in solids of  10$^{-4}${\it M$_{*}$} (Section \ref{solids}) and all this solid material accreted into larger bodies (Section \ref{debris}). 
(3) 50\% of the stars are binary, with 20\% of the single A--K2 type stars and 3\% of the single M-dwarfs harboring planets (Section \ref{planets}),
able to eject most of the planetesimals out of the system (Section \ref{ejection}). 
 
The results are listed in Table 1, adding up to a total mass density of interstellar material of $m_{total}$ = 2.2$\cdot$10$^{-7}$ \msol/pc$^{3}$ = 4.5$\cdot$10$^{26}$ g/pc$^{3}$. 

\begin{deluxetable}{lllll}
\tablewidth{0pc}
\tablecaption{Expected Mass Density of Extra-Solar Material}
\tablehead{
\colhead{SpType} &
\colhead{Mass Range} &
\colhead{$\int$ $\xi$({\it M$_{*}$})10$^{-4}${\it M$_{*}$}d{\it M$_{*}$}} &
\colhead{Fraction of} &
\colhead{Mass Density} \\
\colhead{} &
\colhead{(\msol)} &
\colhead{(\msol/pc$^{3}$)} &
\colhead{Contributing Stars} &
\colhead{(\msol/pc$^{3}$)}
}
\startdata
M				&	0.1--0.5		&	2.1$\cdot$10$^{-6}$		&	0.03$\times$0.5		&	3.1$\cdot$10$^{-8}$ \\
MK2			&	0.5--0.8		&	6.1$\cdot$10$^{-7}$		&	0.03$\times$0.5		& 9.1$\cdot$10$^{-9}$ \\
K2G			&	0.8--1		&	4.3$\cdot$10$^{-7}$		&	0.2$\times$0.5		& 4.3$\cdot$10$^{-8}$ \\
GF			&	1--1.8		&	9.1$\cdot$10$^{-7}$		&	0.2$\times$0.5		& 9.1$\cdot$10$^{-8}$ \\
A				&	1.8-2.9\tablenotemark{a}		&	5.1$\cdot$10$^{-7}$		&	0.2$\times$0.5		& 5.1$\cdot$10$^{-8}$ \\
AFGKM 	&	0.1--2.9		&										&	 							& 2.2$\cdot$10$^{-7}$ \\
\enddata
\tablenotetext{a}{We do not consider larger stellar masses because for $M_* >$3 \msol~the snow
line move quickly beyond 10--15 AU before protoplanets form, limiting the formation of gas giant planets.}
\end{deluxetable}

\section{EXPECTED NUMBER DENSITY OF EXTRA-SOLAR PLANETESIMALS WITH $r > R$}
\label{size_dist}

From the total mass density of extra-solar material ($m_{total}$) derived above, we can estimate the number of planetesimals 
formed within a particular size bin by assuming a planetesimal bulk density of 1.5 g/cm$^3$ 
(between that of pure ice and the density inferred for Pluto) and a size distribution.
For the latter, we use the range of size distributions found for the small body population in the solar system as a guide of the 
ranges that may be found in the planetesimals ejected by other systems: 

\begin{itemize}
\item {\it Asteroids.}\\
In the asteroid belt, the current size distribution is well known down to a size of 1 km, showing a wavy
structure with enhancements at 3--4 km and 100 km, and a power-law index for the differential size distribution 
ranging from 2.25 to 3.8. Because this size distribution has evolved with time, more relevant 
is the size distribution at the time when the asteroids were ejected out of the system, i.e., at the time the solar 
system likely contributed to the population of extra-solar planetesimals. Using a collisional evolution model together 
with observational constraints, Bottke et al. (2005) concluded that the size distribution of this ``primordial'' asteroid 
belt followed a broken power law: 
$n(r) \propto r^{-q_1}~{\rm if}~r<r_b$ and 
$n(r) \propto r^{-q_2}~{\rm if}~r>r_b$, 
where {\it r} is the planetesimal radius, {\it r$_b$} $\approx$ 50 km, {\it q$_1$} $\approx$ 1.2 and {\it q$_2$} $\approx$ 4.5. 
This distribution was established early on as a result of a period of collisional activity before 
Jupiter formed (few Myr), and a period of collisional activity  triggered by the planetary embryos (10--100 Myr).

\item {\it Kuiper Belt Objects.}\\
The differential size distribution expected for KBOs, from coagulation models that take into account the collisional 
evolution due to self-stirring, follows a broken power law with
{\it n(r) $\propto$ r$^{-q_1}$} if {\it r $\leq$ r$_1$},  
{\it n(r) $\propto$ constant} if {\it r$_1$ $\leq$ r $<$ r$_0$}, and
{\it n(r) $\propto$ r$^{-q_2}$} if {\it r $\geq$ r$_0$}, 
where {\it r} is the planetesimal radius and $q_1$ $\approx$ 3.5 (resulting from the collisional cascade); 
for a fragmentation parameter, {\it Q$_b$} $\gtrsim$ 10$^5$ erg/g, 
$q_2$ $\approx$ 2.7--3.3, and  {\it r$_0$  $\approx$ r$_1$ $\approx$} 1 km; 
while for {\it Q$_b$} $\lesssim$ 10$^3$ erg/g, $q_2$ $\approx$ 3.5--4, {\it r$_1$  $\approx$} 0.1 km, and
{\it r$_0$  $\approx$} 10--20 km (see review in Kenyon et al. 2008). Recent KBO surveys reviewed by Kenyon et al. (2008) yield $q_2$ $\approx$ 3.5--4 and {\it r$_{max}$} $\sim$ 300--500 km for the cold classical KB 
({\it a} = 42--48 AU, {\it perihelion} $>$ 37 AU, {\it i} $\lesssim$ 4$^\circ$); 
$q_2$ $\approx$ 3 and {\it r$_{max}$} $\sim$ 1000 km for the hot classical KB 
({\it i} $\gtrsim$ 10$^\circ$); and 
$q_2$ $\approx$ 3 and {\it r$_{max}$} $\sim$ 1000 km for the resonant population; in all cases, 
the transitional radius is {\it r$_0$  $\approx$} 20--40 km (for albedo $\sim$ 0.04--0.07). 
From a Hubble Space Telescope survey, Bernstein et al. (2004) found {\it q$_1$} = 2.9 and {\it q$_2$} $>$ 5.85 for the classical KB, and 
{\it q$_1$} $<$ 2.8 and {\it q$_2$} = 4.3 for the excited KB, with {\it r$_b$} $\leq$ 50 km in both cases.   
A pencil-beam search for KBOs with Subaru found  {\it r$_b$} $\approx$ 45$\pm$15($p$/0.04)$^{-0.5}$ km (where $p$ is the albedo -- 
Fuentes, George \& Holman 2009). 
Finally, the most recent Subaru survey by Fraser and Kavelaars (2009), sensitive to KBOs with $r >$ 10 km, found 
{\it q$_1$} = 1.9, {\it q$_2$} = 4.8, and {\it r$_b$} $\leq$ 25--47 km (assuming a 6\% albedo). 

Kenyon et al. (2008) argued that the comparison between the observational and the modeling results indicates that self-stirring 
alone cannot account for the observed size distribution in the KB and that dynamical perturbations 
have played a major role, suggesting that the  size distribution in the KB was frozen after a major 
event of dynamical ejection. Fraser and Kavelaars (2009) argued that the large {\it r$_b$} implies increased collisional evolution, 
while the large {\it q$_2$} suggests there was an early end of the accretion stage. Even though it is tempting to adopt the size
distribution of the KBOs as a proxy of extra-solar planetesimals, the above indicates that the size distribution 
depends significantly on the dynamical/collisional evolution of the population at the time of the ejection. 

\item {\it Comets.}\\
The size distribution of {\it elliptic comets} follows a power law of index $q \approx$ 2.9 for $r >$ 1.6 km, and if including cometary near-earth objects (thought to be extinct elliptic comets) $q \approx$ 2.6.  However, elliptic comets are likely collisional fragments of KBOs and therefore their size distribution is heavily evolved due to collisions and perihelion passages.  More representative of the population of planetesimals ejected by the early solar system might be the less processed {\it nearly isotropic comets}, in particular those in their first perihelion passage. These comets are thought to originate from the OC that was populated in the early solar system by planetesimals in the vicinity of the giant planets. Their size distribution is not known because of the small number of detections, although there are indications that it is shallower than the elliptic comets (see review by Lamy et al. 2004). 
\end{itemize}

Based on the size distributions discussed above, and because it is not known the degree of dynamical/collisional evolution of the extra-solar planetesimals at the time they are ejected from other planetary systems, we consider a size distribution following a broken power-law with these range of parameters: 
\begin{equation}
\begin{split}
n(r) \propto r^{-q_1}~{\rm if}~r<r_b~~~~~~~~~~~~~~~~~~~\\
n(r) \propto r^{-q_2}~{\rm if}~r>r_b~~~~~~~~~~~~~~~~~~~\\
q_1 = {\rm 2.0,~2.5,~3.0,~3.5} ~~~~~~~~~~~~~~~~~\\
q_2 = {\rm 3,~3.5,~4,~4.5,~5}~~~~~~~~~~~~~~~~~~\\
r_b = {\rm 3~km, 30~km,~90~km}~~~~~~~~~~~~~\\
r_{max} \approx~{\rm1000~km~and}~r_{min} \approx~{\rm1}~\mu{\rm m}
\end{split}
\label{eq_size_dist_r_range}
\end{equation}
(For simplicity, we use $r_{min}$ = 1 $\mu$m for all stellar types, although for A-type stars the blow-out radius is $\sim$10 $\mu$m.)

Using the size distributions above, the expected number density of extra-solar planetesimals with radius  $r>R$~($< r_b$) is 
\begin{equation}
\begin{split}
{N_{r>R} = {9(4\pi\rho)^{-1}m_{total} \over {3 \over 4-q_2}[({r_{max} \over r_{b}})^{4-q_2}-1]+{3 \over 4-q_1}[1-({r_{min} \over r_{b}})^{4-q_1}]}}\\
\times \left( {1 \over 1-q_1}[r_b^{1-q_1}-R^{1-q_1}]r_b^{q_1-4}+{1 \over 1-q_2}[r_{max}^{1-q_2}-r_b^{1-q_2}]r_b^{q_2-4}\right).
\end{split}
\label{N_r_gt_R_a}
\end{equation}
(see the Appendix A for details). 
%valid in the case of a broken power-law for the size distribution with $R < r_b$. 

We now use Equation \eqref{N_r_gt_R_a} to calculate the total number density of planetesimals per pc$^3$ with radius $r > R$ for the range of size distributions discussed above (in Equation \eqref{eq_size_dist_r_range}). The results are shown in Figure 1 and Table 2, indicating that the number of planetesimals per pc$^3$ with $r >$ 1 km is in the range O(10$^6$)--O(10$^{10}$), differing very significantly from the value used by McGlynn \& Chapman (1989) of 10$^{13}$ pc$^{-3}$. Using the latter number density, and including corrections for the observability 
of extra-solar comets, McGlynn \& Chapman (1989) concluded that there is a missing  extra-solar comet problem because a significant number of extra-solar 
comets should have already been detected. However, our estimate of the number density of planetesimals is several orders of magnitude smaller so it is not surprising that extra-solar comets have not been detected so far. 

\begin{figure}
\begin{center}
\includegraphics[scale=0.7,angle=90]{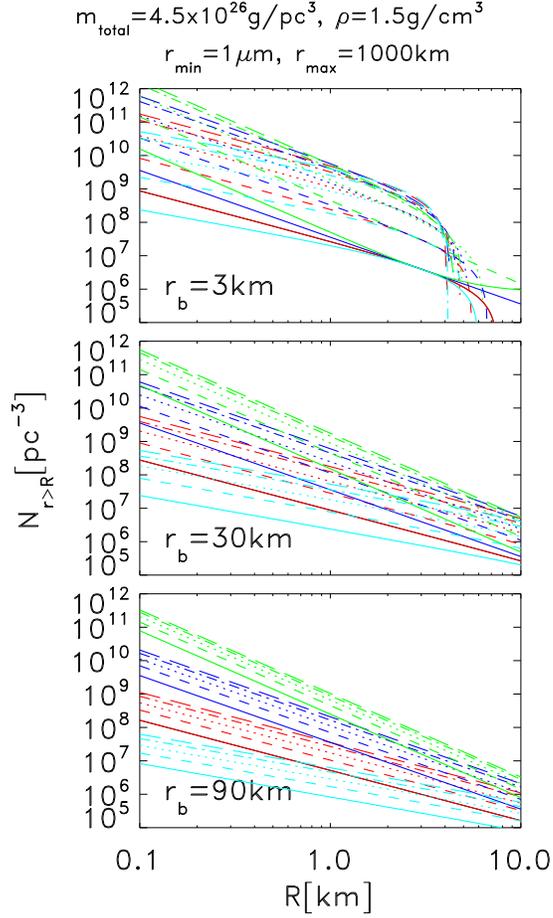}
\end{center}
\caption{Total number density of planetesimals per pc$^3$ with radius $r>R$ calculated from Equation \eqref{N_r_gt_R} for the range of size distributions in  Equation \eqref{eq_size_dist_r_range}. The colors and line types correspond to $q_1$ = 2.0 (light blue), 2.5 (red), 3.0 (blue), and 3.5 (green); 
$q_2$ = 3 (solid), 3.5 (dashed), 4 (dotted), 4.5 (dash-dotted), 5 (long dashed). The different panels correspond to different break radius: 
$r_b$ = 3 km (top panel), $r_b$ = 30 km (middle panel), and $r_b$ = 90 km ({\it bottom panel}). In all cases, the total mass density of extra-solar material  is $m_{total}$ = 2.2$\cdot$10$^{-7}$ \msol/pc$^{3}$ = 4.5$\cdot$10$^{26}$ g/pc$^{3}$ (derived in Section \ref{mass_density}),  the planetesimal bulk density is $\rho$ = 1.5 g/cm$^3$,  {\it r$_{max}$} = 1000 km, and {\it r$_{min}$} = 1 $\mu$m. 
}
\label{nR_1_1000}
\end{figure}

\begin{deluxetable}{lllll}
\tablewidth{0pc}
\tablecaption{Expected $N(r >$ 1 km) and $N(m<$ 24.5)\tablenotemark{a}}
\tablehead{
\colhead{r$_b$} &
\colhead{r$_{min}$} &
\colhead{r$_{max}$} &
\colhead{$N(r >$ 1 km)} &
\colhead{$N(m <$ 24.5)}\\
\colhead{} &
\colhead{} &
\colhead{(km)} &
\colhead{(pc$^{-3}$)} &
\colhead{(deg$^{-2}$)}
}
\startdata

3 km 	& 1 $\mu$m 	& 1000  	& 2$\cdot$10$^7$--5.4$\cdot$10$^9$    		& 2.6$\cdot$10$^{-10}$--2.4$\cdot$10$^{-8}$\\
	 		& 1 $\mu$m 	&  500  	& 4$\cdot$10$^7$--5.4$\cdot$10$^9$ 			& 5.1$\cdot$10$^{-10}$--2.4$\cdot$10$^{-8}$ \\
			& 1 $\mu$m 	&  50  		& 4.1$\cdot$10$^8$--5.5$\cdot$10$^9$ 		& 5.3$\cdot$10$^{-9}$--2.4$\cdot$10$^{-8}$ \\
			& 1 $\mu$m 	&  5  		& 5.3$\cdot$10$^9$--8$\cdot$10$^9$ 			& 3.7$\cdot$10$^{-8}$--3.4$\cdot$10$^{-8}$ \\
	 		& 1 km 			&  1000 	& 2$\cdot$10$^7$--8.8$\cdot$10$^9$ 			& 2.6$\cdot$10$^{-10}$--3.3$\cdot$10$^{-8}$ \\
	 		& 1 km 			&  50  		& 4.1$\cdot$10$^8$--9$\cdot$10$^9$ 			& 5.3$\cdot$10$^{-9}$--3.4$\cdot$10$^{-8}$ \\
	 		& 1 km 			&  5  		& 5.5$\cdot$10$^9$--1.3$\cdot$10$^{10}$ 	& 3.9$\cdot$10$^{-8}$--7.7$\cdot$10$^{-8}$ \\		
		
30 km 	& 1 $\mu$m 	& 1000  	& 2.4$\cdot$10$^6$--1.8$\cdot$10$^9$ 		& 1.1$\cdot$10$^{-10}$--8.3$\cdot$10$^{-9}$\\
	 		& 1 $\mu$m 	&  500 	& 4.8$\cdot$10$^6$--1.8$\cdot$10$^9$ 		& 2.3$\cdot$10$^{-10}$--8.4$\cdot$10$^{-9}$ \\
			& 1 $\mu$m 	&  50  		& 6.7$\cdot$10$^7$--2.2$\cdot$10$^9$ 		& 2.3$\cdot$10$^{-9}$--1.0$\cdot$10$^{-8}$ \\
%			& 100 m 		& 1000  	& 2.6$\cdot$10$^5$--2.0$\cdot$10$^8$ 		& 1.2$\cdot$10$^{-11}$--9.3$\cdot$10$^{-10}$ \\
	 		& 1 km 			&  1000  	& 2.4$\cdot$10$^6$--2$\cdot$10$^9$ 			& 1.1$\cdot$10$^{-10}$--9.5$\cdot$10$^{-9}$ \\
	 		& 1 km 			&  50  		& 6.7$\cdot$10$^7$--2.6$\cdot$10$^9$	 		& 2.3$\cdot$10$^{-9}$--1.2$\cdot$10$^{-8}$ \\		

90 km 	& 1 $\mu$m 	& 1000  	& 8.2$\cdot$10$^5$--1$\cdot$10$^9$ 			& 7.1$\cdot$10$^{-11}$--5.0$\cdot$10$^{-9}$ \\
	 		& 1 $\mu$m 	&  500 	& 1.7$\cdot$10$^6$--1$\cdot$10$^9$ 			& 1.5$\cdot$10$^{-10}$ --5.1$\cdot$10$^{-9}$ \\
%	 		& 100 m 		& 1000  	& 8.9$\cdot$10$^5$--1.1$\cdot$10$^8$ 		& 7.7$\cdot$10$^{-12}$--5.5$\cdot$10$^{-10}$ \\
	 		& 1 km 			& 1000  	& 8.2$\cdot$10$^5$--1.1$\cdot$10$^9$ 		& 7.1$\cdot$10$^{-11}$--5.4$\cdot$10$^{-9}$ \\

\enddata
\tablenotetext{a}{
$N(r >$ 1 km) is the expected number of planetesimals  per pc$^{3}$ with radius $r >$ 1km.  
$N(m <$ 24.5) is the expected number of planetesimals per deg$^2$ with magnitude brighter than 24.5 (the limiting LSST magnitude for a single visit). 
These values are calculated using 
Equation \eqref{N_r_gt_R} for $N(r >$ 1 km)  and 
Equations \eqref{N_m}, \eqref{b_1}, \eqref{b_2}, \eqref{b_3}, and \eqref{A} for $N(m <$ 24.5), for the range of size distributions discussed in Section \ref{size_dist}: $n(r) \propto r^{-q_1}$ for $r < r_b$ and $n(r)  \propto r^{-q_2}$ for $r > r_b$, with $q_1$ = 2.0--3.5, $q_2$ = 3--5, and $r_b$ = 3km, 30 km, and 90 km. 
The total mass density of extra-solar material  is $m_{total}$ = 2.2$\cdot$10$^{-7}$ \msol/pc$^{3}$ = 4.5$\cdot$10$^{26}$ g/pc$^{3}$ (derived in Section \ref{mass_density}), the planetesimal bulk density is $\rho$ = 1.5 g cm$^{-3}$ and the albedo is 6\% (corresponding to inactive comets). 
}
\end{deluxetable}

\section{EXPECTED LUMINOSITY OF EXTRA-SOLAR PLANETESIMALS}
\label{luminosity}

Using the number density of planetesimals in interstellar space derived in Section \ref{size_dist}, we can calculate the distribution of luminosities as observed from the Earth (see details of the calculation in the Appendix B). This will allow us to estimate how many of these objects might be detected, given the limiting magnitude of the survey under consideration. We assume the planetesimals are distributed isotropically and that their size-magnitude of brightness relation is  
\begin{equation}
m = K + 2.5log_{10}(a^{-4}D^{-2}), 
\label{m}
\end{equation}
where $D$ is the planetesimal diameter in km, $a$ is the heliocentric distance\footnote{The distance to the Earth is approximated as the heliocentric distance.} in AU, and $K$ = 18.4 mag, which assumes that all objects have an albedo of 6\%. This albedo is typical of inactive comets, therefore, this size-magnitude of brightness relation would only be valid for inactive planetesimals (i.e., with no comae). Following Fraser \& Kavelaars (2009), the cumulative luminosity function, i.e., the number of objects per deg$^2$ brighter than a given magnitude ($m < m_{max}$) is given by
\begin{equation}
N(m<m_{max}) \simeq b_1 10^{{q_1-1 \over 5} m_{max}} + b_2 10^{{q_2-1 \over 5} m_{max}} + b_3,  
\label{N_m_a}
\end{equation}
with,  
\begin{equation}
\begin{split}
b_1 = A {D_b^{q_1-q_2} 10^{{-K(q_1-1) \over 5}} \over (q_1-1)(5-2q_1)}\left[(D_b^{1/2}10^{(m_{max}-K)/10})^{5-2q_1}-a_1^{5-2q_1}\right], 
\label{b_1_a}
\end{split}
\end{equation}

\begin{equation}
\begin{split}
b_2 = A {10^{{-K(q_2-1) \over 5}} \over (q_2-1)(5-2q_2)}\left[a_2^{5-2q_2}-(D_b^{1/2}10^{(m_{max}-K)/10})^{5-2q_2}\right], 
\end{split}
\label{b_2_a}
\end{equation}

\begin{equation}
\begin{split}
b_3 = A D_b^{1-q_2} {q_1-q_2 \over 3(q_2-1)(q_1-1)}\left[(D_b^{1/2}10^{(m_{max}-K)/10})^3-a_1^3\right], 
\end{split}
\label{b_3_a}
\end{equation}

\begin{equation}
\begin{split}
A =   \left({1.5\cdot10^{13} \over 3\cdot10^{18}}\right)^3 \left({\pi \over 180}\right)^2\left(2\cdot10^{-5}\right)^3
{9(4\pi\rho)^{-1}m_{total}\left(D_b\right)^{q_2-4} \over {3 \over 4-q_2}[({D_{max} \over D_{b}})^{4-q_2}-1]+{3 \over 4-q_1}[1-({D_{min} \over D_{b}})^{4-q_1}]}, 
\label{A_a}
\end{split}
\end{equation}
where the planetesimal diameter $D$ is in km, its heliocentric distance $a$ is in AU, its bulk density $\rho$ is in g/cm$^3$, and the total mass density of planetesimals $m_{total}$ is in g/pc$^3$. Equations \eqref{b_1_a}, \eqref{b_2_a}, \eqref{b_3_a}, and \eqref{A_a} are only valid for $q_1 \neq$ 1, $q_1 \neq$ 2.5, $q_2 \neq$ 1, and  $q_2 \neq$ 2.5.  

Figure 2 shows the number of planetesimals brighter than a given magnitude ($m<m_{max}$) per deg$^2$ calculated from Equations \eqref{N_m_a}, \eqref{b_1_a}, \eqref{b_2_a}, \eqref{b_3_a}, and \eqref{A_a}, with the parameter ranges in Equation \eqref {eq_size_dist_r_range}. These are  the cumulative luminosity distributions that would correspond to the cumulative size distributions in Figure 1. As mentioned above, we assume the planetesimals have an albedo of 6\%, i.e., they are inactive. To fulfill this constrain we consider planetesimals at heliocentric distances beyond 5 AU, i.e., $a_1$ = 5 AU and $a_2 \rightarrow \infty$.
Table 2 shows the expected number of planetesimals per deg$^2$ with $m <$ 24.5 (the limiting magnitude of LSST for a single visit),  for different values of the minimum and maximum planetesimal diameter ($D_{min}$ and $D_{max}$).

\begin{figure}
\begin{center}
\includegraphics[scale=0.7,angle=90]{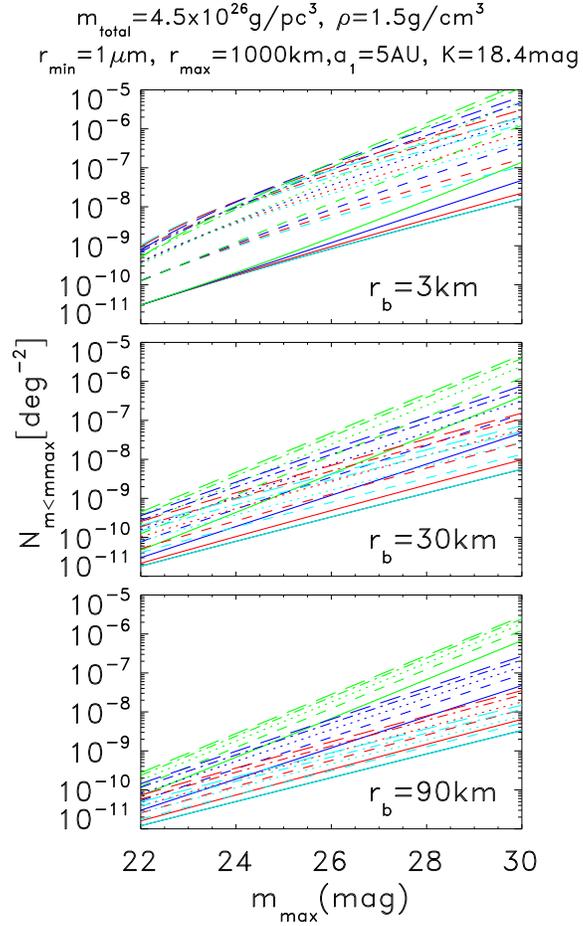}
\end{center}
\caption{Number of planetesimals brighter than a given magnitude ($m<m_{max}$) per deg$^2$  for the size distributions in Figure 1 (same color code).  We assume the planetesimals have an albedo of 6\%. 
}
\label{nR_1_1000}
\end{figure}

\section{DISCUSSION}
\label{dicussion}

LSST is expected to achieve a depth of 24.5 magnitudes for each single visit. Table 2 shows that for the broad range of size distributions considered, the expected number of extra-solar planetesimals  per deg$^2$ brighter than this limiting magnitude is O(10$^{-10}$)--O(10$^{-8}$). In the total area that will be covered  by LSST, 2$\cdot$10$^4$ deg$^2$,  we would expect to find O(10$^{-6}$)--O(10$^{-4}$) extra-solar planetesimals.  But we need to consider the following: 
\begin{itemize} 
\item LSST will observe the sky 1000 times\footnote{Because the extra-solar planetesimals are moving objects, we are limited by the single visit magnitude and in this case the number of visits does not increase the depth.} in 10 years. Considering that the velocity of the extra-solar planetesimals  with respect to the Sun is approximately the velocity of the Sun with respect to the LSR, $v_{LSR} \sim$16.5 \kms, we expect the number of extra-solar planetesimals inside 5 AU to "refresh" approximately once per year.  The numbers in Table 2 were calculated assuming albedos of 6\%, corresponding to inactive comets, and to be consistent with this the objects are assumed to be at distances beyond 5 AU.  This means that the "refreshing" of the population of extra-solar planetesimals within the 10 year lifetime of the LSST survey increases the estimated number of detections  by {\it less} than a factor of 10. 
\item The gravitational focusing factor at a heliocentric distance $R$ is $G_f = \left(1+{4G M_{\odot} \over  v^2 R}\right)^{1/2}$, so this effect can increase the number of expected detections by a factor of $\sim G_f^3$. For $R$ =  5 AU,  $G_f  \sim$ 2; because the objects are at distances beyond 5 AU, gravitational focusing could increase the number of detections by {\it less} than a factor of 10. 
\end{itemize}
The above two considerations lead to an increase in the expected number of detections of {\it less} than a factor of 100, i.e. we expect LSST to find at most O(10$^{-4}$)--O(10$^{-2}$) extra-solar planetesimals\footnote{The upper limit could be 10 times larger if we were to assume that all the solids in the protoplanetary disks accrete into objects with  a very narrow range of  sizes, from 1 to 5 km with a break radius of $r_{b}$ = 3 km, but this very narrow size distribution is not supported by observations of small bodies in the solar system, nor by core accretion models of planet formation that predict the formation of larger bodies.} during the duration of its lifetime.

\begin{figure}
\begin{center}
\includegraphics[scale=0.7,angle=90]{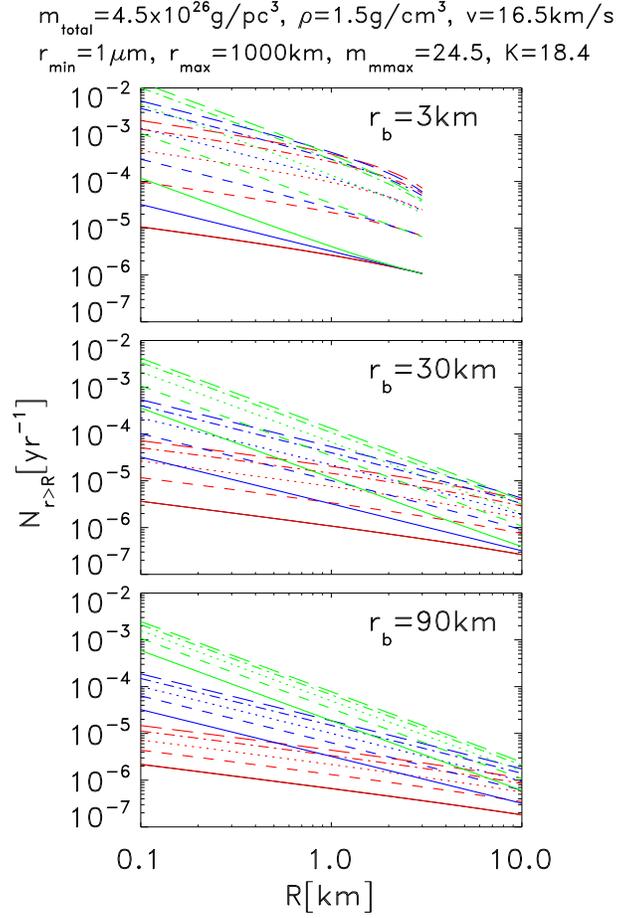}
\end{center}
\caption{Expected number of incoming detectable extra-solar planetesimals larger than radius $R$, per year, for the size distributions in Figure 1 (same color code).  We assume the planetesimals have an albedo of 6\%. 
}
\label{figure_flux}
\end{figure}

The effect of the refreshing of the extra-solar planetesimal population mentioned above can be calculated explicitly by estimating the expected incoming flux of extra-solar planetesimals larger than a given radius $R$. This flux is $\int\limits_{R}^{r_{max}} n(r)\sigma(r)vdr$, where $v$ is the planetesimal velocity with respect to the Sun ($v \sim v_{LSR} \sim$16.5 \kms), $n(r)$ is the planetesimal differential number density (from Equations \eqref{eq_size_dist_r} and \eqref{A1p_A2p_detailed}), and  $\sigma$ is a "cross-section", $\sigma(r) = \pi a^2$. In this lalter expression, $a(r)$ is the maximum distance at which a planetesimal of radius $r$ could be detected,  $a(r) = (2r)^{1/2}10^{(m_{max}-K)/10}$, where $m_{max}$ is the limiting detectable magnitude, $r$ is the planetesimal radius in km and $K$ = 18.4 (assuming an albedo of 6\% -- see Section \ref{luminosity} and Equation \eqref{m}).  For $m_{max}$ = 24.5 mag (limiting magnitude of LSST for a single visit), and using 
Equations \eqref{eq_size_dist_r} and \eqref{A1p_A2p_detailed}, we get that the number of incoming detectable extra-solar planetesimals larger than radius $R$, per year, is 

\begin{equation}
\int\limits_{R}^{r_{max}} n(r)\sigma(r)vdr = 
4.5\cdot10^{-19}\left[A'_1\left({r_b^{-q_1+2}\over-q_1+2} - {R^{-q_1+2}\over-q_1+2}\right) + A'_2\left({r_{max}^{-q_2+2}\over-q_2+2} - {r_{b}^{-q_2+2}\over-q_2+2}\right)\right],
\label{flux}
\end{equation}
where $A'_1$ and $A'_2$ are given by Equation  \eqref{A1p_A2p_detailed} (with $\rho$ = 1.5 g/pc$^3$, $m_{total}$ = 4.5$\cdot10^{26}$ g/pc$^3$ -- see \S \ref{mass_density}), and $R$, $r_{b}$ and $r_{max}$ are in cm. The results, shown in Figure \ref{figure_flux}, indicate that the likelihood of detection
is larger for small objects close to the Sun, compared to larger bodies at
a wider range of distances; the faintness (and hence smaller $\sigma$) of
small objects is compensated by their larger number density.

We conclude that the fact that not a single extra-solar comet has been detected so far is not in contradiction with current  knowledge (as pointed out by Jewitt 2003), and that the probability that LSST will detect a single $inactive$ extra-solar comet during the duration of its lifetime is very small ($\sim$ 0.01\%--1\%): a non-detection of inactive comets is expected from our current knowledge of the distribution of stellar masses,  the typical  protoplanetary disk mass (assuming opacities typical of mm-sized particles\footnote{If larger particles were found to be common in the protoplanetary disks, the disk masses would increase affecting the estimated number of extra-solar comets.}),  the frequency of planetesimals and planets, and the distribution of sizes of small  bodies. Note, however, that the likelihood of detection is significantly larger for small objects close to the Sun. At these small heliocentric distances, the expected number of small detectable objects in Figure \ref{figure_flux} is underestimated because of increased gravitational focusing, and more importantly, because these objects might have become active and more easily detectable. The study of how cometary activity affects the expected number of detections is left for future work.

\subsection{Caveats}
In our calculation we have assumed that extra-solar planetesimals have albedos of 6\%, similar to inactive comets. To be consistent with this assumption we have considered extra-solar planetesimals located beyond 5 AU, i.e., before comets start outgassing. However, the properties of extra-solar planetesimals are not known. In the solar system, albedos can vary from 2\% up to 70\%. For example, Centaurs have no comae, either because they have already lost their volatiles, or because their surfaces consist of non-volatile complex organics and silicate mantles processed by the bombardment of energetic particles  (Jewitt 2004). Extra-solar planetesimals have long interstellar travel times and is expected that they are exposed to cosmic rays for long periods of time. It is therefore reasonable to assume that their crusts are highly processed and, as a consequence, have low albedos before they become active --  as we have assumed in our calculations. 

But we do not  know at what heliocentric distance extra-solar planetesimals activate and how bright they might become because the composition of the ices could be very different from solar system comets; for example, CO, if dominant, could enhance the comet activity far from the star. New nearly isotropic comets (i.e., those in their first perihelion passage) are anomalously bright, probably because of their high volatile content (accreted from the interstellar medium), and/or the blow-off of their  cosmic-ray-processed  crust, and/or the amorphous-to-crystalline H$_2$O ice phase transition (occurring at 5 AU during their first perihelion passage -- Dones et al. 2004).  Extra-solar comets might also become unusually bright when they become active and their highly processed crust is blown off.  This makes the estimate of how many $active$ extra-solar comets might be detected very challenging. 

Another issue to keep in mind is that extra-solar planetesimals may break easily. 
From modeling and observations of inactive nearly isotropic comets, it is found that 99\% become unobservable as they evolve inward from the OC, probably due to breaking into smaller pieces; as these comets become inactive, they only have a 1\% chance of becoming dormant, the other 99\% must have disrupted; this 99\% comes from comparing the number of returning comets to those predicted based on the number of new nearly isotropic comets (Levison et al. 2002; Dones et al. 2004). For comparison, Jupiter family comets do not seem to break so easily and about 60\% become dormant,\footnote {Possible reasons  are because they contain less volatiles (due to being stored at higher temperatures), and/or they are more porous (because of more frequent collisions) and therefore less susceptible to disruption from the build-up of volatiles, and/or they are subject to smaller temperature gradients (Levison et al. 2002).} compared to 1\% (Levison et al. 2002). If extra-solar comets are similar to the unaltered nearly isotropic (OC) comets,  it may be possible that a high fraction of them break up and become unobservable before we have the chance to detect them. 

\begin{center}
{\bf APPENDIX A}
\end{center}
\begin{center}
{\bf DERIVATION OF THE NUMBER DENSITY}
\end{center}
Using the size distributions in Equation \eqref{eq_size_dist_r_range}, we calculate the expected number density of extra-solar planetesimals with radius $r>R$.  We adopt a broken power law for the mass distribution,\footnote{We start the discussion in terms of the mass distribution (rather than the size distribution) because we use the total mass density of extra-solar material ($m_{total}$ calculated in Section \ref{mass_density}) to normalize the distribution.}, 
\begin{equation}
\begin{split}
n(m) = A_1m^{-\alpha_1}~~{\rm for}~m < m_b\\
n(m) = A_2m^{-\alpha_2}~~{\rm for}~m > m_b,
\end{split}
\label{eq_size_dist_m}
\end{equation} 
where $A_1$ and $A_2$ are calculated from 
continuity, $A_1m_b^{-\alpha_1} = A_2m_b^{-\alpha_2}$, 
and from normalizing to the expected mass density over the full mass range, 
$m_{total} = \int\limits_{m_{min}}^{m_{b}}A_1m^{-\alpha_1}mdm+ \int\limits_{m_{b}}^{m_{max}}A_2m^{-\alpha_2}mdm$, yielding
\begin{equation}
\begin{split}
{A_1 = {m_{total} \over 
m_b^{-\alpha_1+\alpha_2} {m_{max}^{-\alpha_2+2}-m_{b}^{-\alpha_2+2} \over -\alpha_2+2}+
{m_{b}^{-\alpha_1+2}-m_{min}^{-\alpha_1+2} \over -\alpha_1+2}}}\\
{A_2 = {m_{total} \over 
{m_{max}^{-\alpha_2+2}-m_{b}^{-\alpha_2+2} \over -\alpha_2+2}+
m_b^{-\alpha_2+\alpha_1}{m_{b}^{-\alpha_1+2}-m_{min}^{-\alpha_1+2} \over -\alpha_1+2}}}.
\end{split}
\label{A1_A2}
\end{equation}
Assuming all planetesimals have the same bulk density, $\rho$, the above mass distribution is equivalent to a size distribution of  
\begin{equation}
\begin{split}
n(r) = A'_1r^{-q_1}~~{\rm for}~r < r_b\\ 
n(r) = A'_2r^{-q_2}~~{\rm for}~r > r_b,
\end{split}
\label{eq_size_dist_r}
\end{equation} 
where $A'_1$, $A'_2$, $q_1$ and $q_2$ are derived from 
$n(m)dm=n(r)dr$ and given by, 
\begin{equation}
\begin{split}
{A'_1 = 3^{\alpha_1}(4\pi\rho)^{-\alpha_1+1}A_1},\\
{A'_2= 3^{\alpha_2}(4\pi\rho)^{-\alpha_2+1}A_2}, 
\end{split}
\label{A1p_A2p}
\end{equation}
with $q_1 = 3\alpha_1-2$ and $q_2 = 3\alpha_2-2$.
Substituting the expressions in Equation \eqref{A1_A2} into \eqref{A1p_A2p}, and using $m_i = {4 \over 3}\pi\rho r_i^3$ and $\alpha_i = { q_i+2 \over 3}$, we get
\begin{equation}
\begin{split}
{A'_1 = {9(4\pi\rho)^{-1}m_{total}r_b^{q_1-4} \over {3 \over 4-q_2}[({r_{max} \over r_{b}})^{4-q_2}-1]+{3 \over 4-q_1}[1-({r_{min} \over r_{b}})^{4-q_1}]}}\\
{A'_2 = {9(4\pi\rho)^{-1}m_{total}r_b^{q_2-4} \over {3 \over 4-q_2}[({r_{max} \over r_{b}})^{4-q_2}-1]+{3 \over 4-q_1}[1-({r_{min} \over r_{b}})^{4-q_1}]}}.
\end{split}
\label{A1p_A2p_detailed}
\end{equation}
The total number density of planetesimals with radius $r>R$~($< r_b$)  is given by $N_{r>R} = \int\limits_{R}^{r_{b}}A'_1r^{-q_1}dr+ \int\limits_{r_{b}}^{r_{max}}A'_2r^{-q_2}dr$, and from the expressions in \eqref{A1p_A2p_detailed} we get,
\begin{equation}
\begin{split}
{N_{r>R} = {9(4\pi\rho)^{-1}m_{total} \over {3 \over 4-q_2}[({r_{max} \over r_{b}})^{4-q_2}-1]+{3 \over 4-q_1}[1-({r_{min} \over r_{b}})^{4-q_1}]}}\\
\times \left( {1 \over 1-q_1}[r_b^{1-q_1}-R^{1-q_1}]r_b^{q_1-4}+{1 \over 1-q_2}[r_{max}^{1-q_2}-r_b^{1-q_2}]r_b^{q_2-4}\right).
\end{split}
\label{N_r_gt_R}
\end{equation}
%valid in the case of a broken power-law for the size distribution with $R < r_b$. 

\begin{center}
{\bf APPENDIX B}
\end{center}
\begin{center}
{\bf DERIVATION OF THE LUMINOSITY FUNCTION}
\end{center}
The number of planetesimals with radius $r$ to $r+dr$ (with $r$ in cm) per pc$^3$ is given by Equations \eqref{eq_size_dist_r} and \eqref{A1p_A2p_detailed}, with the possible range of parameters listed in \eqref{eq_size_dist_r_range}.  If we assume the planetesimals are distributed isotropically,  the number of plantesimals with radius $r$ to $r+dr$ (in cm) and heliocentric distance $a$ to $a+da$ (in AU) per deg$^2$ would be given by 
$n(r)dr \left({1.5\cdot10^{13} \over 3\cdot10^{18}}\right)^3 4\pi a^2 da {1 \over 4 \pi \left({180 \over \pi}\right)^2}$, 
where ${1.5\cdot10^{13}  \over 3\cdot10^{18}}$ is the number of pc in an AU and  
$4\pi \left({180 \over \pi}\right)^2$ is the number of deg$^2$ is a sphere. For the derivation of the luminosity function, we need to have the size distribution in terms of the planetesimal diameter  $D$ in km. The transformation is given by $n(r)dr = S(D)dD{1 \over 2\cdot10^{-5}}$; from Equations \eqref{eq_size_dist_r} and \eqref{A1p_A2p_detailed}, we get 
\begin{equation}
\begin{split}
S(D) = A''_1D^{-q_1}~~{\rm for}~D < D_b\\
S(D) = A''_2D^{-q_2}~~{\rm for}~D > D_b
\end{split}
\label{eq_size_dist_D}
\end{equation}
with, 
\begin{equation}
\begin{split}
{A''_1 = {9(4\pi\rho)^{-1}m_{total}\left({D_b \over 2\cdot10^{-5}}\right)^{q_1-4} \over {3 \over 4-q_2}[({D_{max} \over D_{b}})^{4-q_2}-1]+{3 \over 4-q_1}[1-({D_{min} \over D_{b}})^{4-q_1}]} (2\cdot10^{-5})^{q_1}}\\
{A''_2 = {9(4\pi\rho)^{-1}m_{total}\left({D_b \over 2\cdot10^{-5}}\right)^{q_2-4} \over {3 \over 4-q_2}[({D_{max} \over D_{b}})^{4-q_2}-1]+{3 \over 4-q_1}[1-({D_{min} \over D_{b}})^{4-q_1}]} (2\cdot10^{-5})^{q_2}}. 
\label{A1pp_A2pp_detailed}
\end{split}
\end{equation}

The number of objects per deg$^2$, with diameters from $D_{min}$ to $D_{max}$ (in km) and heliocentric distances from $a_1$ to $a_2$ (in AU), is given by 
\begin{equation}
\begin{split}
{N =  
\left({1.5\cdot10^{13} \over 3\cdot10^{18}}\right)^3 \left({\pi \over 180}\right)^2\left({1 \over 2\cdot10^{-5}}\right) } \\
\times {\left[
\int\limits_{a_1}^{a_2}\int\limits_{D_{min}}^{D_{b}}
A''_1D^{-q_1} dD a^2 da + 
\int\limits_{a_1}^{a_2}\int\limits_{D_{b}}^{D_{max}}
A''_2D^{-q_2} dD a^2 da
\right]} \\
{= \left({1.5\cdot10^{13} \over 3\cdot10^{18}}\right)^3 \left({\pi \over 180}\right)^2\left({1 \over 2\cdot10^{-5}}\right)A''_2} \\
\times \left[
\int\limits_{a_1}^{a_b} a^2 \left[{A''_1 \over A''_2}\int\limits_{D_{min}}^{D_{b}}D^{-q_1}dD + 
\int\limits_{D_{b}}^{D_{max}}D^{-q_2}dD
\right]da + 
\int\limits_{a_b}^{a_2} a^2 \int\limits_{D_{min}}^{D_{max}}D^{-q_2}dD 
\right], 
\end{split}
\label{N_D}
\end{equation}
where $a_b$ (in AU) is the maximum distance at which a planetesimal of diameter $D_b$ (in km) can be detected, determined by the limiting magnitude of the survey. 

We now follow Fraser and Kavelaars (2009) to calculate  the luminosity function corresponding to the size distribution in Equation \eqref{N_D}.  As in their calculation, we assume that the size-magnitude of brightness relation is  $m = K + 2.5log_{10}(a^{-4}D^{-2})$, where $D$ is the planetesimal diameter in km, $a$ is the heliocentric distance\footnote{The distance to the Earth is approximated as the heliocentric distance.} in AU, and $K$ = 18.4 mag, which assumes that all objects have an albedo of 6\%. Using the relation above,  the maximum distance at which a planetesimal of diameter $D_b$ (in km) can be detected is  $a_b = D_b^{1/2}10^{(m_{max}-K)/10}$.  Following Fraser \& Kavelaars (2009), the cumulative luminosity function, i.e. the number of objects per deg$^2$ brighter than a given magnitude ($m < m_{max}$) is given by, 
\begin{equation}
N(m<m_{max}) \simeq b_1 10^{{q_1-1 \over 5} m_{max}} + b_2 10^{{q_2-1 \over 5} m_{max}} + b_3,  
\label{N_m}
\end{equation}
with,  
\begin{equation}
\begin{split}
b_1 = A \int\limits_{a_1}^{D_b^{1/2}10^{(m_{max}-K)/10}} a^2 D_b^{q_1-q_2} {a^{2(1-q_1)} \over q_1-1} da 10^{{-K(q_1-1) \over 5}} \\
= A {D_b^{q_1-q_2} 10^{{-K(q_1-1) \over 5}} \over (q_1-1)(5-2q_1)}\left[(D_b^{1/2}10^{(m_{max}-K)/10})^{5-2q_1}-a_1^{5-2q_1}\right], 
\label{b_1}
\end{split}
\end{equation}

\begin{equation}
\begin{split}
b_2 = A \int\limits_{D_b^{1/2}10^{(m_{max}-K)/10}}^{a_2} a^2 {a^{2(1-q_2)} \over q_2-1} da 10^{{-K(q_2-1) \over 5}} \\
= A {10^{{-K(q_2-1) \over 5}} \over (q_2-1)(5-2q_2)}\left[a_2^{5-2q_2}-(D_b^{1/2}10^{(m_{max}-K)/10})^{5-2q_2}\right], 
\end{split}
\label{b_2}
\end{equation}

\begin{equation}
\begin{split}
b_3 = A \int\limits_{a_1}^{D_b^{1/2}10^{(m_{max}-K)/10}} a^2 da D_b^{1-q_2} \left[{q_1-q_2 \over (q_2-1)(q_1-q)}\right] \\
= A D_b^{1-q_2} {q_1-q_2 \over 3(q_2-1)(q_1-1)}\left[(D_b^{1/2}10^{(m_{max}-K)/10})^3-a_1^3\right], 
\end{split}
\label{b_3}
\end{equation}

\begin{equation}
\begin{split}
A =  \left({1.5\cdot10^{13} \over 3\cdot10^{18}}\right)^3 \left({\pi \over 180}\right)^2\left({1 \over 2\cdot10^{-5}}\right)A''_2 \\
= \left({1.5\cdot10^{13} \over 3\cdot10^{18}}\right)^3 \left({\pi \over 180}\right)^2\left(2\cdot10^{-5}\right)^3
{9(4\pi\rho)^{-1}m_{total}\left(D_b\right)^{q_2-4} \over {3 \over 4-q_2}[({D_{max} \over D_{b}})^{4-q_2}-1]+{3 \over 4-q_1}[1-({D_{min} \over D_{b}})^{4-q_1}]}, 
\label{A}
\end{split}
\end{equation}
where the planetesimal diameter $D$ is in km, its heliocentric distance $a$ is in AU, its bulk density $\rho$ is in g/cm$^3$ and the total mass density of planetesimals $m_{total}$ is in g/pc$^3$. Equations \eqref{b_1}, \eqref{b_2}, \eqref{b_3} and \eqref{A} are only valid for $q_1 \neq$ 1, $q_1 \neq$ 2.5, $q_2 \neq$ 1, and  $q_2 \neq$ 2.5.  

\begin{center} {\it Acknowledgments} \end{center}
We are grateful to Matt Holman for insightful discussion and Darin Ragozzine for useful comments. A.M.M. acknowledges funding via  the Ram\'on y Cajal Fellowship from the  Spanish Government, the Michelson Fellowship  from JPL, the Lyman Spitzer Fellowship from Princeton University and  the Spitzer archival grant 40412. She thanks the Institute of Advance Studies for hospitality during the summer 2008. E.L.T. was supported in part by the World Premier International Research Center Initiative (WPI Initiative), MEXT, Japan.

\end{document}